\documentclass[a4paper,oneside,final,notitlepage,onecolumn,12pt]{article}
\usepackage{amsfonts}
\usepackage{epsf}
\usepackage{graphicx}% Include figure files
\usepackage{amssymb,eso-pic}
\usepackage{latexsym}
\usepackage{tabularx}
\usepackage{amsxtra,accents}
\usepackage{hyperref}
\usepackage{t1enc}
\usepackage{amsmath}
\usepackage[T1]{fontenc}
\usepackage{amsmath}
\usepackage{authblk}

\usepackage{framed}
\usepackage{enumerate}
\usepackage{bm} 
\usepackage{bbm}
\usepackage{mathrsfs}
\usepackage{amsthm}
\usepackage{latexsym}
\usepackage{tikz}
\usepackage{mathtools}
\DeclareGraphicsExtensions{.jpg,.jpeg,.pdf,.png,.eps}

\usepackage{dutchcal} % \mathcal: also small letters. \mathbcal = bold ones. 

\usepackage{ifpdf,epsfig,array,amsmath,amssymb}

\usepackage{psfrag}
\psfrag{na}{\footnotesize $n^a$}   
\psfrag{ta}{\footnotesize $\tau^a$}   
\psfrag{vna}{\footnotesize $\check{n}^a$}
\psfrag{vNa}{\footnotesize $\check{N}^a$}
\psfrag{scrW}{\footnotesize $\mathscr{W}_{\rho_0}$}
\psfrag{t=t1}{\footnotesize $\Sigma_{\tau_1}$}
\psfrag{t=t2}{\footnotesize $\Sigma_{\tau_2}$}
\psfrag{r=all}{\footnotesize $\rho=const$}
\psfrag{hna}{\footnotesize $\widehat{n}^a$}
\psfrag{tna}{\footnotesize $\tilde{n}^a$}
\psfrag{hab,Kab}{\footnotesize $h_{ab}, K_{ab}$}
\psfrag{thab,tKab}{\footnotesize $\tilde h_{ab}, \tilde K_{ab}$}
\psfrag{hhab,hKab}{\footnotesize $\widehat h_{ab}, \widehat K_{ab}$}
\psfrag{chab,cKab}{\footnotesize $\check{h}_{ab}, \check{K}_{ab}$}

\newcommand{\interior}[1]{\accentset{\smash{\raisebox{-0.12ex}{$\scriptstyle\circ$}}}{#1}\rule{0pt}{-.3ex}}
\newcommand{\indot}[1]{\accentset{\smash{\raisebox{-0.1ex}{$\scriptstyle{\bigcdot}$}}}{#1}\rule{0pt}{2.3ex}}
\makeatletter
\newcommand*{\bigcdot}{}% Check if undefined
\DeclareRobustCommand*{\bigcdot}{%
	\mathbin{\mathpalette\bigcdot@{}}%
}
\newcommand*{\bigcdot@scalefactor}{.5}
\newcommand*{\bigcdot@widthfactor}{1.15}
\newcommand*{\bigcdot@}[2]{%
	\sbox0{$#1\vcenter{}$}% math axis
	\sbox2{$#1\cdot\m@th$}%
	\hbox to \bigcdot@widthfactor\wd2{%
		\hfil
		\raise\ht0\hbox{%
			\scalebox{\bigcdot@scalefactor}{%
				\lower\ht0\hbox{$#1\bullet\m@th$}%
			}%
		}%
		\hfil
	}%
}
\makeatother
\definecolor{DGREEN}{rgb}{0,0.65,0.65}
\definecolor{grey1}{rgb}{0.52, 0.52, 0.51}

\usepackage[normalem]{ulem}
\usepackage{color}
\definecolor{blue}{rgb}{0,0,1}
\definecolor{red}{rgb}{1,0,0}

\makeatother

\DeclareFontFamily{OT1}{rsfs}{} \DeclareFontShape{OT1}{rsfs}{m}{n}{
<-7> rsfs5 <7-10> rsfs7 <10-> rsfs10}{}
\DeclareMathAlphabet{\mathscr}{OT1}{rsfs}{m}{n}

\def\sc{{\hskip 3.5pt {{}^{{}^{{}_{{}_{\bowtie}}}}} \kern -8.pt{}}}  
\def\SC{{\hskip 3.5pt {{}^{{}^{{}^{{}_{{}_{\bowtie}}}}}} \kern -10.5pt{}}}

\newtheorem*{example*}{Example}%[section]
\newtheorem*{condition*}{Condition}%[section]
\usepackage[normalem]{ulem}
\usepackage{color}

\newcounter{mnotecount}%[section]

\newcommand{\mnotex}[1]%{}
{\protect{\stepcounter{mnotecount}}$^{\mbox{\footnotesize $\bullet$\themnotecount}}$ 
	\marginpar{\color{red}
		\raggedright\tiny\em
		$\!\!\!\!\!\!\,\bullet$\themnotecount: #1} }

\begin{document}

\title{Quasilocal spin-angular momentum and the construction of axial vector fields} 
%\title{On axial vector fields and spin-angular momentum}

\author{Istv\'an R\'acz \footnote{E-mail address: {\tt racz.istvan@wigner.hu}}}

\affil{HUN-REN Wigner RCP, H-1121 Budapest, Konkoly Thege Mikl\'{o}s \'{u}t  29-33, Hungary}

\maketitle

\begin{abstract}
A novel procedure is presented that allows the construction of all axial vector fields on Riemannian two-spheres. Using these axial vector fields and the centre-of-mass unit sphere reference systems, a constructive definition of quasilocal spin-angular momentum is introduced. Balance relations are also derived, with respect to arbitrary Lie-propagated unit sphere reference systems, to characterize the angular momentum transports in spacetimes without symmetries.
\end{abstract}
 
%"The whole problem with the world is that fools and fanatics are always so certain of themselves, but wiser men so full of doubts" Bertrand Russell

%%%%%%%%%%%%%%%%%%%% INTRODUCTION %%%%%%%%%%%%%%%%%%%%%%%%%%%%%%%%%%%%%%

\section{Preliminaries}\label{preliminaries}
\setcounter{equation}{0}
 
It is widely believed that comprehensive theoretical studies of truly dynamical processes are not possible without the use of appropriate quasilocal quantities such as mass, energy, linear and angular momentum (see \cite{Hawking-1968, Geroch, Penrose-1982a, Christodoulou-Yau-1986}). However, since general relativity is a metric theory of gravity, the definition of such quantities remains challenging. This scenario is further complicated by the desire for simplicity in determining these quantities and the demand for practicality. Despite the controversies surrounding the notions of energy and linear momentum, there seems to be an unexpected consensus on the form  of the quasilocal spin angular momentum (see \cite{Szabados-2004,Brown-York-1993, Ashtekar-Krishnan-2002, Booth-Fairhurst-2005, Eric-2005, Hayward-2006, Korzyński-2007} and references therein). It is assumed that if we knew all the ``axial'' vector fields tangential to the boundary of a finite domain, then the angular momentum carried by that domain could have been determined. However, despite considerable efforts, the determination of these axial vector fields has remained a challenge for decades.

\medskip 

quasilocal quantities are assigned to finite spatial regions bounded by compact orientable two-surfaces. It is widely accepted that if a meaningful quasilocal angular momentum can be defined, it must be given as a surface integral over a (sufficiently smooth) topological two-sphere $\mathscr{S}$. The most commonly used proposal is based on the use of the connection one-form, given as \cite{Szabados-2004, Brown-York-1993, Ashtekar-Krishnan-2002, Booth-Fairhurst-2005, Eric-2005, Hayward-2006, Korzyński-2007}\,\footnote{Other definitions of quasilocal or global angular momentum have been proposed in the literature, see \cite{Szabados-2004} for a comprehensive review, and also \cite{Penrose-1982, Vickers-1983, Kijowski-2001, Yau-2016, Klainerman-2022} for alternative proposals.}
\begin{equation}
	\omega_a = - \mathcal{k}_b\nabla_a \ell^b\,,
\end{equation}
where $\mathcal{k}^a$ and $\ell^a$ are future-pointing null normals to $\mathscr{S}$, scaled such that $\mathcal{k}^a \ell_a=-1$, while  $\nabla_a$ stands for covariant derivative operator compatible with the metric $g_{ab}$ of an ambient spacetime. 
The quasilocal spin-angular momentum, with respect to an axial vector field $\phi^a$ that is tangent to $\mathscr{S}$, is given by
\begin{equation}\label{eq: qlam-gen}
	J[\phi] = - (8\, \pi)^{-1}\int_{\mathscr{S}} \phi^a\omega_a\, \widehat{{\boldsymbol{\epsilon}}} \,,
\end{equation}   
where ${\widehat{\boldsymbol{\epsilon}}}$ stands for the volume element of the metric $\widehat{\gamma}_{ab}$ induced by the spacetime metric on $\mathscr{S}$.  By this definition, only the part $\widehat{\omega}_a=\widehat{\gamma}_a{}^b\,\omega_b$ of $\omega_a$ projected onto $\mathscr{S}$ counts in the above integral. Note also that the existence of axial vector field(s) is a matter of critical importance in the above definition. Such a vector field has to be tangent to $\mathscr{S}$, and it should also satisfy the following criteria 
\begin{itemize}
\item[(i)] it has closed orbits (with period $2\,\pi$, and with two poles), and
\item[(ii)] it has vanishing divergence with respect to the induced connection $\widehat{D}_a$. 
\end{itemize}  
Condition $(i)$ ensures that the integral curves of an axial vector field smoothly foliate $\mathscr{S}$, except at its poles, by circles. Condition $(ii)$, on the other hand, guarantees the invariance of the volume form $\widehat{\boldsymbol{\epsilon}}$, under the flow generated by $\phi^a$, as then  $\mathscr{L}_\phi\,\widehat{{\boldsymbol{\epsilon}}}=\widehat{D}_a\phi^a\,\widehat{{\boldsymbol{\epsilon}}}=0$, which always holds for Killing vector fields, although in general an axial vector field need not be a Killing vector field on $\mathscr{S}$.

\medskip 

As we proceed, note first that the future-pointing null normals $\mathcal{k}^a$ and $\ell^a$ are not unique on $\mathscr{S}$, since they can be rescaled by using an arbitrary smooth real function $f$ on $\mathscr{S}$ such that for $\mathcal{k}'{}^a= e^f \, \mathcal{k}^a$ and $\ell'{}^a=e^{-f}\, \ell{}^a$ the scaling relation $\mathcal{k}'{}^a \ell'_a=-1$ remains intact \cite{Booth-Fairhurst-2005}. However, under such a rescaling, the connection one-form changes as $\omega'_a =\omega_a -\widehat{D}_a f$. This leads to 
\begin{equation}\label{eq: qlam-g0}
	\phi^a\omega'_a = \phi^a[\omega_a-\widehat{D}_a f] = \phi^a\omega_a - \widehat{D}_a (f\phi^a) + f (\widehat{D}_a \phi^a) \,,
\end{equation}
which, together with the vanishing of $\widehat{D}_a\phi^a$, guarantees that $J[\phi]$, given in \eqref{eq: qlam-gen}, is gauge invariant, i.e., its value is insensitive to this type of rescalings.

\medskip 

In light of the construction recalled above, it is also worth considering the following alternative formulation. Suppose $\mathscr{S}$ is embedded in a spacelike hypersurface $\Sigma$ in an ambient spacetime $(M,g_{ab})$. Let $n^a$ and $\widehat{n}^a$ be unit vector fields on  $\mathscr{S}$ such that they are both orthogonal to $\mathscr{S}$, and also orthogonal and parallel to $\Sigma$, respectively. Assume also that $n^a$ is future-directed. The combinations $\mathcal{k}^a=(n^a-\widehat{n}^a)/\sqrt2$ and $\ell^a=(n^a+\widehat{n}^a)/\sqrt2$ are then properly scaled future directed null normals. We also have 
\begin{equation}
	\omega_a =-\tfrac12\,(n_b-\widehat{n}_b) \nabla_a (n^b+\widehat{n}^b) = -\tfrac12\,(n_b\nabla_a\widehat{n}^b - \widehat{n}_b \nabla_a n^b) = \widehat{n}^b \nabla_a n_b \,,
\end{equation}
which implies 
\begin{equation}
	\widehat{\omega}_a =\widehat{\gamma}_a{}^e \widehat{n}^b \nabla_e n_b = \widehat{\gamma}_a{}^e K_{eb}\widehat{n}^b={\rm \bf k}_a\,,
\end{equation}
where $K_{ab}$ and ${\rm \bf k}_a$ denote the extrinsic curvature of $\Sigma$ and its vector projection, respectively. Note that since $\phi^a$ is tangent to $\mathscr{S}$ while $\widehat{n}^a$ is normal to it, we also have that $\phi^a{\rm \bf k}_a = \phi^a(K_{ab}-\tfrac12\,h_{ab}K) \,\widehat{n}^b$, which implies that 
\begin{equation}\label{eq: qlam-gen2}
	J[\phi] = - (8\, \pi)^{-1}\int_{\mathscr{S}} \phi^a(K_{ab}-\tfrac12\,h_{ab}K) \,\widehat{n}^b \, \widehat{{\boldsymbol{\epsilon}}} = - (8\, \pi)^{-1}\int_{\mathscr{S}} \phi^a{ \rm \bf k}_a \, \widehat{{\boldsymbol{\epsilon}}} \,.
\end{equation}
The integral expression in the middle also verifies the equivalence of the quasilocal angular momentum expression proposed by Brown and York (given in \cite{Brown-York-1993} under the assumption that $\phi^a$ is an axial Killing vector field) and that given by \eqref{eq: qlam-gen} with respect to an axial vector field $\phi^a$. This also suggests that, if the axial vector field, $\phi^a$, tends to an asymptotic Killing vector field, then the quasilocal angular momentum, given by either \eqref{eq: qlam-gen} or \eqref{eq: qlam-gen2}, should also tend to the global ADM angular momentum. This is because the first integral expression in \eqref{eq: qlam-gen2} is exactly of the form used in the definition of the corresponding ADM charge.

\medskip

A few comments are in order. Obviously, the unit vector field $n^a$ is well defined in a neighborhood of $\mathscr{S}$ in $\Sigma$. Note, however, that the extendibility of $\widehat{n}^a$ within $\Sigma$ was also implicitly assumed. It can be checked that the integral expression is insensitive to the applied extension. Moreover, there are infinitely many spacelike hypersurfaces through $\mathscr{S}$, and the associated ambiguity in the unit vector fields $n^a$ and $\widehat{n}^a$ corresponds to the rescaling ambiguity in the future-pointing null normals $\mathcal{k}^a$ and $\ell^a$ discussed above. Following an analogous argument, the vanishing of the divergence $\widehat{D}_a\phi^a$ guarantees the invariance of the resulting quasilocal angular momentum.   
 
\section{Construction of axial vector fields}\label{sec: axial-fields} 
\setcounter{equation}{0}  

First, assume that some spherical coordinates $(\vartheta,\varphi)$ and the unit sphere metric $ds^2=d\vartheta^2 +\sin\vartheta^2\,d\varphi^2$, also denoted by $\interior{\gamma}_{ab}$, have been chosen on $\mathscr{S}$. Note that the Poincaré uniformization theorem always guarantees the existence of such spherical coordinates on a Riemannian two-sphere.\,\footnote{By referring to the diffeomorphism $\Psi:\mathscr{S} \to \mathbb{S}^2$, discussed in section \ref{sec: quasilocal angular momentum}, such coordinates can be defined by the inverse $\Psi^{-1}$ of $\Psi$ which takes the spherical coordinates $(\vartheta,\varphi)$ from $\mathbb{S}^2$ to $\mathscr{S}$, while also taking the unit sphere metric $\interior{\gamma}_{ab}$ to $(\Psi^{-1})^*\interior{\gamma}_{ab}$. By slightly abusing this notation, in this section, we will simply write $\interior{\gamma}_{ab}$ instead of $(\Psi^{-1})^*\interior{\gamma}_{ab}$.} In fact, it is not the existence of such spherical coordinates but the abundance of them that causes problems. Nevertheless, we suppress this ambiguity for the moment, to which we will return in section \ref{sec: quasilocal angular momentum}.

\medskip

Now consider the Riemannian metric $\widehat{\gamma}_{ab}$ given on $\mathscr{S}$. , Given also a choice of spherical coordinates and an associated unit sphere metric $\interior{\gamma}_{ab}$ on $\mathscr{S}$ our goal is to construct all the axial vector fields relevant for $(\mathscr{S},\widehat{\gamma}_{ab})$. To do this, consider an arbitrary sufficiently smooth vector field $\chi^a$ on $\mathscr{S}$ and recall that
\begin{equation}\label{eq: questionable}
	\widehat{D}_a\chi^a = \mathbb{D}_a\chi^a + C^a{}_{ae}\chi^e\,,
\end{equation}
where $\mathbb{D}_a$ is the $\interior{\gamma}_{ab}$ compatible covariant derivative operator, and $C^c{}_{ab}$ is the $(1,2)$-type tensor field relating the action of $\widehat{D}_a$ and $\mathbb{D}_a$. Accordingly, we have 
\begin{equation}\label{C-tensor}
	C^a{}_{ab} = \tfrac12\,\widehat{\gamma}{}^{ae}\mathbb{D}_b \widehat{\gamma}{}_{ae} = \tfrac12\, \widehat{\gamma}{}^{ae}(\partial_b \widehat{\gamma}{}_{ae}-\interior{\Gamma}^f{}_{ba}\widehat{\gamma}{}_{fe}-\interior{\Gamma}^f{}_{be}\widehat{\gamma}{}_{af}) = \tfrac12\, \widehat{\gamma}{}^{ae}\partial_b \widehat{\gamma}{}_{ae} - \interior{\Gamma}^a{}_{ab}
	\,,
\end{equation}
where the relation $\widehat{\gamma}{}^{ae}\widehat{\gamma}{}_{af}=\interior{\gamma}{}^{ae}\interior{\gamma}{}_{af}=\delta^e{}_f$ was used.
Then, using the Jacobi identity of the matrix calculus, we get
\begin{equation}\label{C-tensor2}
	C^a{}_{ab} = \partial_b \ln \sqrt{{\widehat{\gamma}}\,\,} - \partial_b \ln \sqrt{{\interior{\gamma}}\hskip-0.1cm\phantom{)}} = \sqrt{{\interior{\gamma}}/{\widehat{\gamma}}\,}\,\, \partial_b\big[\sqrt{{\widehat{\gamma}}/{\interior{\gamma}}\,}\big] \,,
\end{equation}
where $\widehat{\gamma}$ and $\interior{\gamma}$ denote the determinants of the metrics $\widehat{\gamma}_{ab}$ and $\interior{\gamma}_{ab}$ on $\mathscr{S}$, respectively. This, together with \eqref{eq: questionable} and \eqref{C-tensor}, yields
\begin{equation}
	\widehat{D}_a \chi^a = \mathbb{D}_a \chi^a + \sqrt{{\interior{\gamma}}/{\widehat{\gamma}}\,}\, \big(\chi^a\,\mathbb{D}_a\big[\sqrt{{\widehat{\gamma}}/{\interior{\gamma}}\, }\big]\big) = \sqrt{{\interior{\gamma}}/{\widehat{\gamma}}\,}\,\,\mathbb{D}_a \big[\sqrt{{\widehat{\gamma}}/{\interior{\gamma}}\,}\,\chi^a\big] \,. 
\end{equation}
The last relation then implies that if $\interior{\phi}{}^a$ is an axial Killing vector field with respect to the unit two-sphere metric $\interior{\gamma}_{ab}$, then $\widetilde{\phi}{}^a=\sqrt{{\interior{\gamma}} /{\widehat{\gamma}}\,}\, \interior{\phi}{}^a$ is `almost' an axial vector field with respect to the metric $\widehat{\gamma}{}_{ab}$ on $\mathscr{S}$. Note that $\widetilde{\phi}{}^a$ is divergence-free by construction. It also has closed orbits and admits two poles, but its integral curves are not yet periodic with $2\,\pi$. To get the desired axial vector field ${\phi}{}^a$ on $\mathscr{S}$, we have to rescale $\widetilde{\phi}{}^a$ by applying the averaging factor 
\begin{equation}\label{eq: averaging-factor}
	\underline{\sqrt{{\widehat{\gamma}}/{\interior{\gamma}}}}_{\,[{\tiny{\interior{\phi}}}]} = \frac1{2\,\pi} \int_{0}^{2\,\pi}\sqrt{{\widehat{\gamma}}/{\interior{\gamma}}}  \,\,d\interior{\varphi}\,,
\end{equation}
which is constant along the integral curves of the axial Killing vector field $\interior{\phi}{}^a$, where $\interior{\varphi}$ denotes the corresponding $2\,\pi$-periodic axial coordinate. Note that the vector field 
\begin{equation}\label{eq: axial}
	{\phi}{}^a = \underline{\sqrt{{\widehat{\gamma}}/{\interior{\gamma}}}}_{\,[{\tiny{\interior{\phi}}}]}\, \left[\sqrt{{\interior{\gamma}}/{\widehat{\gamma}}\,}\, \interior{\phi}{}^a\right]
\end{equation}
on $\mathscr{S}$, though it depends on $\interior{\phi}{}^a$ in a nonlocal way, satisfies both conditions $(i)$ and $(ii)$, thus it is an axial vector field of the Riemannian two-sphere $(\mathscr{S},\widehat{\gamma}{}_{ab})$. To see that under this rescaling condition $(ii)$ remains intact, recall that the average $\underline{\sqrt{{\widehat{\gamma}}/{\interior{\gamma}}}}_{\,[{\tiny{\interior{\phi}}}]}$ is constant along the integral curves of $\widetilde{\phi}{}^a$, hence the divergence $\widehat{D}_a{\phi}{}^a = \underline{\sqrt{{\widehat{\gamma}}/{\interior{\gamma}}}}_{\,[{\tiny{\interior{\phi}}}]}\, \big(\widehat{D}_a\widetilde{\phi}{}^a\big) + \widetilde{\phi}{}^a\widehat{D}_a\underline{\sqrt{{\widehat{\gamma}}/{\interior{\gamma}}}}_{\,[{\tiny{\interior{\phi}}}]}$ is zero throughout $\mathscr{S}$. Note also that if $\interior{\phi}{}^a$ happens to be an axial Killing vector field with respect to the metric $\widehat{\gamma}{}_{ab}$, then both factors in \eqref{eq: axial} are constant along the integral curves of $\interior{\phi}{}^a={(\partial_{\interior{\varphi}})}^a$, so they compensate each other, implying, as expected, ${\phi}{}^a=\interior{\phi}{}^a$. 

\medskip

We have seen that an axial Killing vector field of a unit sphere metric on $\mathscr{S}$ can always be used to construct an axial vector field with respect to the metric $\widehat{\gamma}{}_{ab}$ on $\mathscr{S}$. It is of critical importance to know if there can exist an axial vector field $\phi^a$ on $(\mathscr{S},\widehat{\gamma}{}_{ab})$ which is not of the form \eqref{eq: axial}. To see that such an axial vector field cannot exist recall that $\phi^a$ as an axial vector field must have closed orbits, with period $2\,\pi$, and it must admit two poles. Label the closed orbits between the two poles with $\vartheta$, taking values from the interval $(0,\pi)$, and rescale $\phi^a$ along the closed orbits by the reciprocal of the averaging factor in \eqref{eq: averaging-factor} with consistently replacing ${\interior{\gamma}}$ by $\sin^2\vartheta$. This gives a $\varphi$ coordinate along the closed orbits, for which they are periodic with $2\,\pi$.  Moreover, with these $\vartheta$ and $\varphi$ we can define the unit sphere metric by $ds^2=d\vartheta^2 +\sin\vartheta^2\, d\varphi^2$, so that the axial vector field $\phi^a$ we started with is related to $\interior{\phi}{}^a=\partial^a_\varphi$ by \eqref{eq: axial}. Obviously, the unit sphere metric just constructed  is not unique. 
In fact, using an alternative labeling of the closed orbits of the axial vector field $\phi^a$ by $\vartheta'=\vartheta'(\vartheta)$, the corresponding azimuthal coordinates also undergo a generic transformation of the form $\varphi'=\varphi'(\vartheta,\varphi)$. Since the unit sphere metrics defined by the coordinates $(\vartheta,\varphi)$ and $(\vartheta',\varphi')$, respectively, must be conformally related to each other, the coordinate transformations $\vartheta'=\vartheta'(\vartheta)$ and $\varphi'=\varphi'(\vartheta,\varphi)$ must also satisfy the requirements formulated by equations (II.6) and (II.7) of \cite{Sachs_1962}. 
	
\medskip

Note that the diffeomorphism invariance of integrals on manifolds guarantees that the relation
\begin{align}\label{eq: JR3}
	J[\phi] =  {}& - (8\, \pi)^{-1}\int_{\mathscr{S}} \phi^a{ \rm \bf k}_a \, \widehat{{\boldsymbol{\epsilon}}} = - (8\, \pi)^{-1}\int_{\mathbb{S}^2} \underline{\sqrt{{\widehat{\gamma}}/{\interior{\gamma}}}}_{\,[{\tiny{\interior{\phi}}}]}\,\,\interior{\phi}{}^a{ \rm \bf k}_a \, \interior{{\boldsymbol{\epsilon}}}\,,
\end{align}
holds for any axial vector field ${\phi}{}^a$ of the form \eqref{eq: axial}, where $\interior{{\boldsymbol{\epsilon}}}$ denotes the volume element of the unit sphere, $\mathbb{S}^2$, and where \eqref{eq: qlam-gen2} and the relation $\sqrt{{\interior{\gamma}}/{\widehat{\gamma}}\,} \, \widehat{{\boldsymbol{\epsilon}}} =  \interior{{\boldsymbol{\epsilon}}}$ were used. As mentioned above, while $J[\phi]$ is a linear functional of ${\phi}{}^a$, by \eqref{eq: JR3} the dependence of $J[\phi]$ on the axial Killing vector field $\interior{\phi}{}^a$ is nonlinear, and more importantly, it is nonlocal.\,\footnote{A completely different approach is used in \cite{Cook-Whiting-2007,Beetle-Wilder-2014}. First, the closest analog of a Killing vector field on $\mathscr{S}$ is determined by minimizing certain functionals. Then, to obtain the quasilocal angular momentum, the true axial Killing vector field is replaced by the approximate one in the Brown-York integral. In contrast, our approach explores the space of axial vector fields on $\mathscr{S}$ and selects a well-defined, compact subset of this space.  We define  the quasilocal spin-angular momentum  by taking the maximum of the integral expression \eqref{eq: JR3} over this compact subset. Exploring the relationship between these approaches, if one exists, is beyond the scope of this paper but would be interesting.}

\medskip

We close this section by pointing out that once spherical coordinates $(\vartheta,\varphi)$ are fixed on $\mathscr{S}$, there is a natural way to assign a three-dimensional Euclidean vector, denoted by  $\vec{J}\,[{\phi}]$, to the domain bounded by $\mathscr{S}$. This is  achieved by referring to the unique constant linear combination $\interior{\phi}^a=\nu_i\,\interior{\phi}_i{}^a$ of the axial Killing vector field $\interior{\phi}^a$ on $(\mathscr{S},\interior{\gamma}{}_{ab})$, where the coefficients $(\nu_1,\nu_2,\nu_3)$ are supposed to form a unit vector in $\mathbb{R}^3$, while
\begin{align}
	\interior{\phi}_1{}^a & = -\sin\varphi \,(\partial_\vartheta)^a - \cot \vartheta\,\cos\varphi \,(\partial_\varphi)^a\,, \label{eq:rotation-unit-1}\\
	\interior{\phi}_2{}^a & = \cos\varphi \,(\partial_\vartheta)^a - \cot \vartheta\,\sin\varphi\,(\partial_\varphi)^a\,, \label{eq:rotation-unit-2}\\	
	\interior{\phi}_3{}^a & = (\partial_\varphi)^a\,,\label{eq:rotation-unit-3}
\end{align}
stand for the three canonical axial Killing vector fields on the unit sphere $({\mathbb{S}}^2,\interior{\gamma}_{ab})$, where ${\mathbb{S}}^2$ is represented by $\{\vec{x}\in \mathbb{R}^3| \, \|\vec{x}\|=1\}$. Once an orthonormal frame $\{\vec{e}_1,\vec{e}_2,\vec{e}_3\}$ of $\mathbb{R}^3$ is chosen, the Euclidean coordinates are defined as $x_i=\vec{x}\,\vec{e}_i$ such that 
\begin{equation}
	\|\vec{x}\|=x_1^2+x_2^2+x_3^2=1\,.
\end{equation}
 It is noteworthy that the axial Killing vector fields $\interior{\phi}_1{}^a,\interior{\phi}_2{}^a,\interior{\phi}_3{}^a$, given by \eqref{eq:rotation-unit-1}-\eqref{eq:rotation-unit-3}, are the generators of rotations about the $x_1,x_2,x_3$-axes of a unit sphere in $\mathbb{R}^3$. Thus, once the spherical coordinates $(\vartheta,\varphi)$ are given on the unit sphere, a quasilocal spin-angle momentum vector $\vec{J}\,[{\phi}]$ can be defined by the constant linear combination $\interior{\phi}^a=\nu_i\,\interior{\phi}_i{}^a$. The absolute value of $J[{\phi}]$, given by the integral expression \eqref{eq: JR3}, can be interpreted as the magnitude, while the unit vector $(\nu_1,\nu_2,\nu_3)$ in $\mathbb{R}^3$ can be interpreted as the direction of $\vec{J}\,[{\phi}]$.

\section{The quasilocal spin-angular momentum}\label{sec: quasilocal angular momentum}

In this section we will explore the freedom we have in choosing spherical coordinates and unit sphere metrics on $\mathscr{S}$. By restricting attention to the centre-of-mass unit sphere reference systems, we fix this freedom, which allows the introduction of the notion of intrinsic quasilocal spin-angular momentum.  

\medskip

The Poincaré uniformization theorem guarantees that for any Riemannian metric ${\widehat{\gamma}}_{ab}$ on $\mathscr{S}$ there always exists a diffeomorphism $\Psi:\mathscr{S} \to \mathbb{S}^2$, a positive real function $\Omega$ and spherical coordinates $(\vartheta,\varphi)$ on $\mathbb{S}^2$ so that the unit sphere metric ${\interior{\gamma}}_{ab}$ takes the form 
\begin{equation}
	ds^2=d\vartheta^2 +\sin\vartheta^2\,d\varphi^2\,,
\end{equation}
and the Riemannian metrics ${\widehat{\gamma}}_{ab}$ and ${\interior{\gamma}}_{ab}$ are conformally related such that the relation  
\begin{equation}\label{eq: Poincare}
	\Psi^*{\widehat{\gamma}}_{ab}=\Omega^2\,{\interior{\gamma}}_{ab}\,.
	%\sqrt{{\widehat{\gamma}}/{\interior{\gamma}}}\cdot{\interior{\gamma}}_{ab}
\end{equation}
holds. Note that we then also have that  $\Omega^2=\sqrt{(\Psi{}^*{\widehat{\gamma}})/{\interior{\gamma}}}$.
It is well known that neither the conformal factor nor the underlying spherical coordinates $(\vartheta,\varphi)$ are unique \cite{Sachs_1962}. It is also known that the stereographic counterparts, $z=\cot(\vartheta/2)\cdot e^{\mathbbm{i}\,\varphi}$ and $z'=\cot(\vartheta'/2)\cdot e^{\mathbbm{i}\,\varphi'}$, of the conformally related spherical coordinates $(\vartheta,\varphi)$ and $(\vartheta',\varphi')$ on $\mathbb{S}^2$, respectively, are related by the M\"obius transformations 
\begin{equation}\label{eq: frac-lin-transf} 
	z'=\frac{a\,z + b}{c\,z + d}\,,% \quad {\rm and} \quad z'=\frac{a\,\bar{z} + b}{c\,\bar{z} + d}\, 
\end{equation} 
where $a,b,c,$ and $d$ are arbitrary complex numbers subject to the condition $ad-bc=1$ \cite{Sachs_1962,Newman-1970}. 
Correspondingly, the composition of two successive (orientation preserving) conformal transformations of the unit sphere onto itself can be represented by the composition of fractional linear transformations of the form \eqref{eq: frac-lin-transf}. Thus the conformal transformations of the unit sphere, $({\mathbb{S}^2},\interior{\gamma}_{ab})$, onto itself can be seen to be isomorphic to the six-(real)-parameter restricted Lorentz group which has $SL(2,\mathbb{C})$ as its double covering \cite{Sachs_1962,Newman-1970,Penrose-Rindler-1987}. 

\medskip 

These finite transformations on $\mathbb{S}^2$ can also be seen to be generated by the three axial Killing vector fields $\interior{\phi}_1{}^a,\interior{\phi}_2{}^a,\interior{\phi}_3{}^a$, given by  \eqref{eq:rotation-unit-1}-\eqref{eq:rotation-unit-3},
together with the three proper conformal Killing vector fields
\begin{align}\label{eq:conformal-unit}
	\interior{\xi}_1{}^a & = -\cos\vartheta\cos\varphi \,(\partial_\vartheta)^a + \frac{\sin\varphi}{\sin\vartheta} \,(\partial_\varphi)^a \\
	\interior{\xi}_2{}^a & = -\cos\vartheta\sin\varphi \,(\partial_\vartheta)^a - \frac{\cos\varphi}{\sin\vartheta}\,(\partial_\varphi)^a \\	
	\interior{\xi}_3{}^a & = \sin\vartheta\,(\partial_\vartheta)^a\,.
\end{align}
These six vector fields are also known to satisfy the commutation relations 
\begin{align}\label{eq: commutation}
	&[\interior{\phi}_i,\interior{\phi}_j]=-\epsilon_{ijk}\,\interior{\phi}_k \\
	&[\interior{\xi}_i,\interior{\xi}_j]=\epsilon_{ijk}\,\interior{\phi}_k \\
	&[\interior{\xi}_i,\interior{\phi}_j]=[\interior{\phi}_i,\interior{\xi}_j]=-\epsilon_{ijk}\,\interior{\xi}_k\,,
\end{align} 
so they form the Lie algebra $so(1,3)$, where $\epsilon_{ijk} $ is the Levi-Civita antisymmetric symbol \cite{Newman-1970} (see also \cite{Korzyński-2007}).

\medskip 

It is known that any linear combination of the form $\nu_i\,\interior{\phi}_i{}^a$, where the coefficients $(\nu_1,\nu_2,\nu_3)$ form a unit vector in $\mathbb{R}^3$, is an axial vector field (in the sense defined above), while no linear combination of the proper conformal Killing vector fields is axial. Moreover, while the orbit of $\nu_i\,\interior{\phi}_i{}^a$ with respect to the adjoint action reads as 
\begin{align}
	\interior{\phi}\,'_i{}^a & = \nu_i\,(\nu_k\interior{\phi}_k{}^a) + \cos\theta \Big[\,\interior{\phi}_i{}^a-\nu_i\,(\nu_k\interior{\phi}_k{}^a)\,\Big] - \sin\theta\, \big(\epsilon_{ijk}\nu_j\interior{\phi}_k{}^a\big)\,,\label{eq: andj-so3-phi}\\
	\interior{\xi}\,'_i{}^a & =  \nu_i\,(\nu_k\interior{\xi}_k{}^a) + \cos\theta \Big[\,\interior{\xi}_i{}^a-\nu_i\,(\nu_k\interior{\xi}_k{}^a)\,\Big] - \sin\theta\, \big(\epsilon_{ijk}\nu_j\interior{\xi}_k{}^a\big)\,,\label{eq: andj-so3-xi}
\end{align}
the integrated action of a proper conformal transformation with respect to  $\nu_i\,\interior{\xi}_i{}^a$ is given by
\begin{align}
	\interior{\phi}\,'_i{}^a & = \nu_i\,(\nu_k\interior{\phi}_k{}^a) + \cosh\theta \Big[\,\interior{\phi}_i{}^a-\nu_i\,(\nu_k\interior{\phi}_k{}^a)\,\Big] - \sinh\theta\, \big(\epsilon_{ijk}\nu_j\interior{\xi}_k{}^a\big)\,,\label{eq: andj-boost-phi}\\
	\interior{\xi}\,'_i{}^a & =  \nu_i\,(\nu_k\interior{\xi}_k{}^a) + \cosh\theta \Big[\,\interior{\xi}_i{}^a-\nu_i\,(\nu_k\interior{\xi}_k{}^a)\,\Big] + \sinh\theta\, \big(\epsilon_{ijk}\nu_j\interior{\phi}_k{}^a\big)\,,\label{eq: andj-boost-xi}
\end{align}
where $\theta$ is the group additive parameter.
These relations clearly show that while $SO(3)$ transformations, see \eqref{eq: andj-so3-phi} and \eqref{eq: andj-so3-xi}, do not mix the generators $\interior{\phi}_i{}^a$ and $\interior{\xi}_i{}^a$, by virtue of \eqref{eq: andj-boost-phi} and \eqref{eq: andj-boost-xi}, the proper conformal transformations do. This simple observation will serve as an important guideline in the following. 

\medskip

As we proceed, first recall that our main goal here is to define the quasilocal spin-angular momentum. In doing so we have to take into account that the group $SL(2,\mathbb{C})$, which was found to be isomorphic to the group of conformal transformations of the unit sphere onto itself, is a noncompact Lie group. In fact, the noncompactness of $SL(2,\mathbb{C})$ is the main obstacle we have to face on the way to define the quasilocal spin-angular momentum. It is also known that if $\Phi$ is an element of the group of conformal transformations of the unit sphere onto itself it admits the decomposition (see, e.g., Lemma 3.4 in \cite{Klainerman-2022})
\begin{equation}
	\Phi=R_1 \cdot B_{N,\tau} \cdot R_2\,,
\end{equation}
where $R_1,R_2$ are ``rotations'' that can be represented by the $SU(2)$ elements
\begin{equation}
	R_i = \left(
	\begin{matrix}
		\alpha_i & \beta_i\\
		-\overline{\beta}_i & \overline{\alpha}_i\\
	\end{matrix}\right)\,,
\end{equation}
with $i=1,2$, while $B_{N,\tau}$ corresponds to the scaling (boost) transformation \cite{Sachs_1962}
\begin{equation}\label{eq: boost-exp-tau}
	\cot(\vartheta'/2)= \tau\cdot\cot(\vartheta/2) \quad {\rm and} \quad \varphi'=\varphi\,,
\end{equation}
with $\tau>0$ and such that $N$ is at the ``north pole'', which can also be represented by the special $SL(2,\mathbb{C})$ matrix
\begin{equation}\label{eq: boost-tau}
	B_{N,\tau} = \left(
	\begin{matrix}
		\sqrt{\tau} & 0\\
		0 & \frac{1}{\sqrt{\tau}}\\
	\end{matrix}\right)\,.
\end{equation}
Note also that $SU(2)$ is the universal cover of $SO(3)$ rotations, and, most importantly, it is a compact subgroup of $SL(2,\mathbb{C})$. 
This observation raises the question whether there is a viable selection rule that leads to the desired subset of conformal transformations. The existence of such a selection rule has been demonstrated by several authors, while focusing on different, usually more complex research topics \cite{Gluck-1975,Onofri-1982,Chang-1987}.\,\footnote{Note that the appendix of \cite{Ashetkar-2022a} also contains an argument proving the existence of a canonical choice for a round metric. Although the reasoning involved is somewhat cumbersome, it does not offer new insights beyond those provided by earlier studies \cite{Gluck-1975,Onofri-1982,Chang-1987, Klainerman-2022}.} For the sake of completeness, we give below a short, self-contained, and simple synthesis of the relevant arguments.

\medskip

Once some spherical coordinates $(\vartheta,\varphi)$ are chosen on $\mathbb{S}^2$ consider the three-vector $\vec{X}$ defined by the integral
\begin{equation}\label{eq: CMUSRS-NO}
	\vec{X}=\int_{{\mathbb{S}^2}} \Omega^2\; \vec{x}\; \interior{{\boldsymbol{\epsilon}}}\,,
	%\vec{X}=\int_{{\mathbb{S}^2}} \sqrt{\widehat{\gamma}/\interior{\gamma}}\; \vec{x}\; \interior{{\boldsymbol{\epsilon}}}\,,
\end{equation}
where $\vec{x}=(\cos\varphi\sin\vartheta,\sin\varphi\sin\vartheta,\cos\vartheta)$. Note that here $\vec{x}$ can be viewed either as a vector pointing from the origin of $\mathbb{R}^3$ to the points of $\mathbb{S}^2$, or as the unit normal vector field on $\mathbb{S}^2$ in $\mathbb{R}^3$. Note also that since $\vec{x}$ is a unit vector in $\mathbb{R}^3$, by virtue of \eqref{eq: Poincare} and \eqref{eq: CMUSRS-NO}, using the relation $\Omega^2\,\interior{{\boldsymbol{\epsilon}}}=\sqrt{(\Psi^*\widehat{\gamma})/\interior{\gamma}}\,\interior{{\boldsymbol{\epsilon}}}=\Psi^*\widehat{{\boldsymbol{\epsilon}}}$ as well, $\vec{X}$ must also be an interior point of the ball of radius $\mathscr{A}=\int_{{\mathbb{S}^2}}\Psi^*\widehat{{\boldsymbol{\epsilon}}}=\int_{{\mathscr{S}}}\widehat{{\boldsymbol{\epsilon}}}$, i.e., $\|\vec{X}\|< \mathscr{A}$. If $\vec{X}$ happens to be the zero vector in $\mathbb{R}^3$ then the underlying $(\vartheta,\varphi)$ coordinates will be referred to as a centre-of-mass unit sphere reference system (CMUSRS).

\medskip

Correspondingly, the desired type of selection rule exists if there exists a conformal diffeomorphism $\Phi$ of the unit sphere onto itself that transforms $\vec{X}$ to the origin in $\mathbb{R}^3$. Recall that a map $\Phi: {\mathbb{S}^2} \to {\mathbb{S}^2}$, with $(\vartheta',\varphi')=\Phi\big((\vartheta,\varphi)\big)$ is a conformal diffeomorphism of the unit sphere onto itself if there exists a positive real function $\omega$ such that both $\Phi^*\big(\interior{\gamma}{}_{ab}{(\vartheta,\varphi)}\big)=\omega^2{(\vartheta',\varphi')} \,\interior{\gamma}{}_{ab}{(\vartheta',\varphi')}$ and $\omega^2=\det(d\Phi)$ holds, where $\det(d\Phi)$ is the Jacobian of the map $\Phi$. Accordingly, our task reduces to finding a conformal diffeomorphism $\Phi: {\mathbb{S}^2} \to {\mathbb{S}^2}$ such that 
\begin{equation}\label{eq: CMUSRS-Phi}
	\int_{{\mathbb{S}^2}} (\Phi^{-1})^*\big(\Omega^2\big)\; \vec{x}\; \interior{{\boldsymbol{\epsilon}}} = \int_{{\mathbb{S}^2}} \big(\Omega^2\,\circ\Phi^{-1}\,\big) \det(d\Phi)\;\vec{x}\; \interior{{\boldsymbol{\epsilon}}} = \int_{{\mathbb{S}^2}} \Omega^2\;(\vec{x}\,\circ\,\Phi)\, \interior{{\boldsymbol{\epsilon}}}=\vec{0}\,.
\end{equation}
where the variable change formula was applied in the second step. Since conformal diffeomorphisms of the unit sphere onto itself are in one-to-one correspondence with the M\"obius transformations, our task reduces to finding an element $g$ in $SL(2,\mathbb{C})$ such that 
\begin{equation}
	{\int_{{\mathbb{S}^2}} \Omega^2\hskip0.02cm(z)\, \vec{x}\hskip0.02cm(g z)\; \interior{{\boldsymbol{\epsilon}}}}=\vec{0}\,.
\end{equation}
To see that such a $g\in SL(2,\mathbb{C})$ really exists, consider the map
\begin{equation}\label{eq: equality}
	\vec{\chi}: SL(2,\mathbb{C}) \to \mathbb{R}^3\,,\quad {\rm with} \quad \vec{\chi}\hskip0.02cm(q)={\int_{{\mathbb{S}^2}} \Omega^2\hskip0.02cm(z)\,\vec{x}\hskip0.02cm(g z)\; \interior{{\boldsymbol{\epsilon}}}}
\end{equation}
which is continuous by construction, maps the unit element $e\in SL(2,\mathbb{C})$ to $\vec{X}=\vec{\chi}(e)$, and, as follows from the above discussion, it maps $SL(2,\mathbb{C})$ to the interior of the ball with radius $\mathscr{A}$ in $\mathbb{R}^3$, i.e., $\|\vec{\chi}\|< \mathscr{A}$.  

\medskip

Now consider the foliation of $SL(2,\mathbb{C})$ by spheres defined as
\begin{equation}\label{eq: Bptau}
	\mathscr{B}_\tau=\{\,B_{p,\tau}\,|\, p\in \mathbb{S}^2 \,\}\,,
\end{equation} 
where $\tau$ takes its value from the interval $1\leq \tau \leq \infty$, while $B_{p,\tau}$ denotes a scale transformation where $p$ now plays the role of the ``north pole'' running through the individual points of the unit sphere $\mathbb{S}^2$. Correspondingly, $B_{p,\tau}$ can also be given by the product
\begin{equation}\label{eq: Bptau2}
	B_{p,\tau}=R_{p\to N}^\dag\cdot B_{N,\tau}\cdot R_{p\to N}\,,
\end{equation}	
where $R_{p\to N}$ is the rotation that takes $p$ to the north pole $N$ and $R_{p\to N}^\dag$ is its adjoint, which is also the inverse of $R_{p\to N}$.  
It follows from \eqref{eq: Bptau2} that the foliation degenerates and gives the unit element $e={\mathscr{B}}_{\tau=1}$ of $SL(2,\mathbb{C})$ for $\tau=1$ and hence $\vec{\chi}\,[\,{\mathscr{B}}_{\tau=1}\,]=\vec{X}$ holds. More importantly, since the unit normal vector field $\vec{x}$ on $\mathbb{S}^2$ is invariant under rotations, thus these transformations do not affect the value of the integral in \eqref{eq: equality}, and since $\lim_{\tau\to\infty}\vec{x}\,(B_{N,\tau}\,z)=\lim_{\tau\to\infty}\vec{x}\circ B_{N,\tau}=(0,0,1)^T$ it also follows that $\lim_{\tau\to\infty}\vec{\chi}\,[\,{\mathscr{B}}_\tau\,]$ gives the sphere with radius $\mathscr{A}$ in $\mathbb{R}^3$, which is actually the boundary of the open ball $\vec{\chi}\,[\,SL(2,\mathbb{C})\,]< \mathscr{A}$. Therefore, by continuity of the map $\vec{\chi}: SL(2,\mathbb{C}) \to \mathbb{R}^3$, we also have that for a sufficiently large $\bar\tau$ value of $\tau$ the sphere $\vec{\chi}\,[\,{\mathscr{B}}_{\bar\tau}\,]$ is close to this boundary in $\mathbb{R}^3$. Since, as noted above, the foliation $\{\,\mathscr{B}_\tau\,|\, 1\leq \tau \leq \infty\,\}$ of $SL(2,\mathbb{C})$ is contractible to the unit element $e$, there must exist $\widetilde{\tau} \in[1,\infty)$ such that $\vec{\chi}\,[\,{\mathscr{B}}_{\widetilde{\tau}}\,]$ contains the origin of $\mathbb{R}^3$.

\medskip

Since $\{\mathscr{B}_\tau\}$ provides a one-parameter foliation of $SL(2,\mathbb{C})$, one may expect that the level set ${\mathscr{B}}_{\widetilde{\tau}}$ is unique up to $SO(3)$ rotations. This kind of uniqueness was verified by Lemma 3.5 in \cite{Klainerman-2022}, which can be reformulated in the notation used here as follows. Suppose that we already have a centre-of-mass unit sphere reference system, i.e., $\vec{X}=\vec{0}$ in \eqref{eq: CMUSRS-NO}. The level set ${\mathscr{B}}_{\widetilde{\tau}}$ can be unique only if for a $B_{N,\tau}$ scaling transformation 
\begin{equation}\label{eq: check-for-SO3}
	 {\int_{{\mathbb{S}^2}} \Omega^2\, \Big[\,\vec{x}\circ B_{N,\tau}-\vec{x}\,\Big]\; \interior{{\boldsymbol{\epsilon}}}}=\vec{0}\,
\end{equation}
holds, then $B_{N,\tau}$ must be trivial, i.e., $\tau=1$ must also hold. To see that this is indeed the case, recall that $\vec{x}=(\cos\varphi\sin\vartheta,\sin\varphi\sin\vartheta,\cos\vartheta)$, which together with the substitutions $\vartheta'= 2\,{\rm arccot}\big(\tau\cdot\cot(\vartheta/2)\big)$ and $\varphi'=\varphi$ [the last two relations follow from \eqref{eq: boost-exp-tau}], can be used to derive
\begin{align}
	x_1\circ B_{N,\tau}-x_1 &= (\tau-1)\,\frac{[\tau\,(1+\cos\vartheta)-(1-\cos\vartheta)]\,\sin\vartheta\cos\varphi}{\tau^2\,(1+\cos\vartheta)+(1-\cos\vartheta)}\\
	x_2\circ B_{N,\tau}-x_2 &= (\tau-1)\,\frac{[\tau\,(1+\cos\vartheta)-(1-\cos\vartheta)]\,\sin\vartheta\sin\varphi}{\tau^2\,(1+\cos\vartheta)+(1-\cos\vartheta)}\\
	x_3\circ B_{N,\tau}-x_3 &= (\tau^2-1)\,\frac{\sin^2\vartheta}{\tau^2\,(1+\cos\vartheta)+(1-\cos\vartheta)}\,. \label{eq: check-for-SO3-third}
\end{align}
Inspecting now the integrand relevant for the third component of \eqref{eq: check-for-SO3}, since, for any $\tau\in [1,\infty)$, all the terms in \eqref{eq: check-for-SO3-third} are non-negative and it is regular, it is straightforward to see that the corresponding integral can only vanish if $\tau=1$. When this happens the integrands of the other two components also vanish automatically which completes the proof of our claim that the centre-of-mass unit sphere reference systems are unique up to $SO(3)$ rotations.

\medskip

By economizing on this uniqueness property of centre-of-mass unit sphere reference systems we can now define the intrinsic spin-angular momentum as follows. Taking into account \eqref{eq: andj-so3-phi} we know that $SO(3)$ always maps the set of axial Killing vector fields of unit sphere reference systems onto itself. Hence we define the intrinsic quasilocal spin-angular momentum as the maximum of the quasilocal integral expressions given by \eqref{eq: JR3} over the compact set of axial Killing vector fields over any of the centre-of-mass unit sphere reference systems
\begin{equation}\label{eq: def-spin-angular-momentum}
	J=\max\Big\{J[\phi] \,\,\big\vert\,\,  \interior{\phi}\,\in \,CMUSRS\,\Big\}\,.
\end{equation}
Note that since the centre-of-mass unit sphere reference systems are related by $SO(3)$ transformations, as is the set of axial Killing vector fields, the maximum is well defined.

\medskip

It is important to have good guesses about the approximate value of $J$, at least in certain special cases. It is worth noting that the closer the metric  $\widehat{\gamma}_{ab}$ is to the round sphere metric on ${\mathscr{S}}$, the closer the value of the average $\underline{\sqrt{{\widehat{\gamma}}/{\interior{\gamma}}}}_{\,[{\tiny{\interior{\phi}}}]}$, relevant for an axial Killing vector field $\interior{\phi}{}^a$, is to the value of the function $\sqrt{\widehat{\gamma}/\interior{\gamma}}$.\,\footnote{By slightly abusing our notation, as we did in section \ref{sec: axial-fields}, in the rest of this paper instead of $\sqrt{{\widehat{\gamma}}/(\Psi^{-1}){}^*{\interior{\gamma}}}$ we will simply write $\sqrt{{\widehat{\gamma}}/{\interior{\gamma}}}$.} In such a special case, the quasilocal spin angular momentum $J$, as defined by \eqref{eq: def-spin-angular-momentum}, can be well approximated as 
\begin{equation}\label{eq: approx}
	J\sim \max\bigg\{- (8\, \pi)^{-1}\int_{\mathscr{S}} \big(\interior{\phi}{}^a{ \rm \bf k}_a\big) \, \widehat{{\boldsymbol{\epsilon}}} \,\,\,\Big\vert\,\,\, \interior{\phi}\,\in \,CMUSRS \,\bigg\}\,,
\end{equation}
where, in the integral, the relation  $\sqrt{\widehat{\gamma}/\interior{\gamma}}\,\interior{{\boldsymbol{\epsilon}}}=\widehat{{\boldsymbol{\epsilon}}}$ was used.

\medskip

As an important special case one may consider the asymptotic limit of the metric $\widehat{\gamma}_{ab}$ which is expected to be a round sphere. Assuming that if the obvious limit is not a centre-of-mass unit sphere reference system it could be transformed into a CMUSRS. Then the approximation in \eqref{eq: approx} gets to be exact in the corresponding limit. Note, however, that it is still necessary to determine the maximum value of the integral expressions $\{- (8\, \pi)^{-1}\int_{\mathscr{S}} \big(\interior{\phi}{}^a{ \rm \bf k}_a\big) \, \widehat{{\boldsymbol{\epsilon}}}\}$, where $\{\interior{\phi}{}^a\}$ is formed by the axial Killing vector fields of $(\mathbb{S}^2,\interior{\gamma}{}_{ab})$. It seems to be plausible that such a limiting procedure will reproduce the global angular momentum charges, regardless of whether the underlying spacelike hypersurfaces tend to spacelike infinity or they tend to a spherical cut of null infinity \cite{Wald-1984,Wald-Zoupas-2000,Ashetkar-2022}. It is also known that Wald and Zoupas \cite{Wald-Zoupas-2000} cleared up the supertranslation ambiguity in the definition of the global Bondi-type angular momentum. It is likely that, in the asymptotic limit, the proposed quasilocal spin-angular momentum possesses the corresponding supertranslation invariance when calculated using a centre-of-mass reference system. It remains to be seen whether these expectations can be verified.

\medskip

It is important to mention that there is at least one certain circumstance where the quasilocal angular momentum expression introduced in this paper turned out to be a very effective tool in analyzing the asymptotic behavior of solutions to the parabolic-hyperbolic form of the constraints in the asymptotically hyperboloidal initial value problem \cite{Csukas-Racz-2025}.

\medskip

Finally, note that by virtue of the discussion in the last paragraph of section \ref{sec: axial-fields}, a quasilocal spin-angular momentum vector field, denoted by $\vec{J}$, can be associated with $J$. In the generic case, both the magnitude and the direction of this vector store important information about the spin-angular momentum content of the region bounded by $\mathscr{S}$. Note that the vectorial nature of the quasilocal spin-angular momentum introduced in this paper fits perfectly with the known vectorial nature of the global angular momentum charges (see, e.g., \cite{Wald-Zoupas-2000,Ashetkar-2022}).

\section{Angular momentum balances}\label{sec: am-balances}

Note that various quasilocal angular momentum balance relations for isolated, trapped, and dynamical horizons have been derived in \cite{Brown-York-1993, Ashtekar-Krishnan-2002, Booth-Fairhurst-2005, Eric-2005, Hayward-2006} simply by substituting some fictitious axial vector fields into \eqref{eq: qlam-gen}.  
It is now tempting to see if meaningful angular momentum balance relations can be derived with a slightly better understanding of the nature of the space of axial vector fields. In this section such balance relations will be derived.

\medskip

As we proceed, note first that no field equations have been imposed so far, so all considerations in the previous sections apply to any four-dimensional metric theory of gravity. In contrast, in the following we will restrict our considerations to spacetimes with a metric satisfying Einstein's equation $G_{ab} = 8\pi\,T_{ab} - \Lambda g_{ab}$, where $T_{ab}$ denotes the energy-momentum tensor and $\Lambda$ is the cosmological constant. Then choose a spacelike hypersurface $\Sigma$ foliated by a one-parameter family of topological two-spheres given by a smooth function $\rho: \Sigma \to \mathbb{R}$ whose $\rho=const$ level sets are individually topological two-spheres, also denoted as $\mathscr{S}_\rho$.\,\footnote{One can obtain such a foliation by simply requiring that, topologically, $\Sigma$ be $\mathscr{S}\times \mathbb{R}$. It may also be of interest to note that the existence of foliations by two-spheres of prescribed mean curvature has been proven in various asymptotically flat and asymptotically hyperbolic settings (see, for example, \cite{Huisken-Yau-1996,Metzger-2007,Huang-2010,Nerz-2018}.} In addition, choose a flow vector field $\rho^a$ on $\Sigma$ that is transverse to the $\rho=const$ foliating two-spheres. Using the one-parameter family of two-spheres $\mathscr{S}_\rho$ on $\Sigma$, synchronized spherical coordinates $(\vartheta,\varphi)$ can then be constructed over $\Sigma$ by Lie-dragging the spherical coordinates $(\vartheta,\varphi)$ given on one of the level sets, say on $\mathscr{S}_{\rho_0}$, along $\rho^a$. Note that the resulting synchronized spherical coordinates $(\vartheta,\varphi)$ are not expected to be centre-of-mass reference systems or to be distinguished in any way.

Note that the kinematic setup outlined above allows us to apply the $2+1$ decomposition of $\Sigma$ using the $\rho=const$ level sets. For example, the vector projection ${\rm \bf k}_a$ of the extrinsic curvature $K_{ab}$ of the spacelike hypersurface $\Sigma$ is well defined on the $\mathscr{S}_\rho$ level sets (for more details, see e.g.\,\cite{Racz_constraints}). As we will see below, the integral expression in \eqref{eq: qlam-gen2}, using the vector projection ${\rm \bf k}_a$ of the extrinsic curvature $K_{ab}$, is the one that best fits the desired balance relation.

\medskip

Now use the synchronized spherical coordinates $(\vartheta,\varphi)$ introduced above and defined in $\Sigma$, together with the associated synchronized unit sphere metric ${\interior{\gamma}}_{ab}$ on the $\mathscr{S}_\rho$ level sets. Note that by construction both of these synchronized structures are invariant with respect to the action of the one-parameter family of diffeomorphisms generated by $\rho^a$. Now choose an axial vector field ${\phi}{}^a$ such that it is of the form \ref{eq: axial}, on the $\mathscr{S}_\rho$ level sets, and that is determined by a $\rho^a$-invariant axial Killing vector field $\interior{\phi}{}^a$ determined by the synchronized unit sphere metric ${\interior{\gamma}}_{ab}$. The variation of $J[\phi]$, with respect to the flow $\rho^a$, can then be written as
\begin{align}\label{eq: varJ}
	\mathscr{L}_{\rho} J[\phi] =  {}& - (8\, \pi)^{-1}\int_{\mathscr{S}_\rho} \mathscr{L}_{\rho}\,\big[\,\phi^a{ \rm \bf k}_a \, \widehat{{\boldsymbol{\epsilon}}}\,\big] = - (8\, \pi)^{-1}\int_{\mathscr{S}_\rho} \mathscr{L}_{\rho}\Big[\,\underline{\sqrt{{\widehat{\gamma}}/{\interior{\gamma}}}}_{\,[{\tiny{\interior{\phi}}}]}\,(\interior{\phi}{}^a{ \rm \bf k}_a \, \interior{{\boldsymbol{\epsilon}}})\,\Big]\,,
\end{align}
where, in the second step, the relation $\sqrt{{\interior{\gamma}}/{\widehat{\gamma}}\,} \, \widehat{{\boldsymbol{\epsilon}}} =  \interior{{\boldsymbol{\epsilon}}}$ was used, and $\mathscr{L}_{\rho}$ denotes the Lie derivative with respect to the flow $\rho^a$. Note that since the coordinates $(\vartheta,\varphi)$ are Lie dragged along $\rho^a$ from $\mathscr{S}_{\rho_0}$ to the other $\rho=const$ level sets, neither the  axial Killing vector field $\interior{\phi}{}^a$, nor the volume element $\interior{{\boldsymbol{\epsilon}}}$ vary with respect to  $\rho^a$. These observations then imply
\begin{align}\label{eq: varJ2}
	\mathscr{L}_{\rho} J[\phi] =  {}&  - (8\, \pi)^{-1}\int_{\mathscr{S}_\rho}\Big[ \big(\mathscr{L}_{\rho}\,\ln\underline{\sqrt{{\widehat{\gamma}}/{\interior{\gamma}}}}_{\,[{\tiny{\interior{\phi}}}]}\,\big)\,({\phi}{}^a{ \rm \bf k}_a) + {\phi}{}^a \big[\mathscr{L}_{\rho}\,{ \rm \bf k}_a\big] \,\Big]\, \widehat{{\boldsymbol{\epsilon}}} \,.	
\end{align}

%\medskip

In evaluating $\mathscr{L}_{\rho}\,{ \rm \bf k}_a$ we can use the vector projection of the momentum constraint formulated by equation (3.1) of \cite{Racz_constraints}. This equation refers to the scalar, $\boldsymbol\kappa=K_{ab}{\widehat{n}}^a{\widehat{n}}^b$, and the trace and trace-free parts ${\rm\bf K}$, $\interior{\rm\bf K}_{ab}$ of the tensor projection ${\rm\bf K}_{ab} = { \widehat \gamma}^{e}{}_{a} { \widehat \gamma}^{f}{}_{b}\,K_{ef}$ of the extrinsic curvature $K_{ab}$ of $\Sigma$, respectively, and also to the lapse, $\widehat{N}$, and the shift, $\widehat{N}{}^a$, of the flow $\rho^a=\widehat{N}{\widehat{n}}^a + \widehat{N}{}^a$,  and to the trace of the extrinsic curvature, $\widehat K$, with respect to the unit normal ${\widehat{n}}^a$ of the $\rho=const$ level sets in $\Sigma$. 
Then the vector projection of the momentum constraint can be seen to have the form 
\begin{align}
	\mathscr{L}_{\rho} {\rm\bf k}{}_a & = \widehat{N}{}^e\widehat D_e {\rm\bf k}{}_{a} + \widehat D_a (\tfrac12\widehat{N}{\rm\bf K}) + \widehat D_a (\widehat{N}\boldsymbol\kappa) - \widehat D^e (\widehat{N}\interior{\rm\bf K}{}_{ea}) - (\widehat{N}{\widehat K})\, {\rm\bf k}{}_{a} + \widehat{N}\,\mathfrak{p}_e\widehat{\gamma}{}^{\,e}{}_{a} \nonumber \\ 
	& = \widehat D_e \big(\widehat{N}{}^e { \rm \bf k}_a - \widehat{N}\interior{\rm\bf K}{}^e{}_{a} \big)+\widehat D_a\big(\widehat{N}\,\big[\tfrac12\,{\rm\bf K} +\boldsymbol\kappa\big] \big) - [\widehat{N}{\widehat K}+\widehat D_e \widehat{N}{}^e] \,{ \rm \bf k}_a + 8\pi\,\widehat{N}\,T_{ef}n^e\widehat{\gamma}{}^{f}{}_{a}\,, \label{par_const_n} 
\end{align}
where the first line of \eqref{par_const_n} differs from (3.1) of \cite{Racz_constraints} only in that the latter is multiplied by the lapse $\widehat{N}$ after decomposing $\mathscr{L}_{\widehat{n}} {\rm\bf k}{}_a$, using $\rho^a=\widehat{N}\widehat{n}{}^a+\widehat{N}{}^a$, and replacing $\indot{n}_a$ with $\indot{n}_a=- \widehat D_a \ln\widehat{N}$. Note also that \eqref{par_const_n} is insensitive to the value of the cosmological constant, since its coefficient $g_{ef}n^e\widehat{\gamma}{}^{f}{}_{a}$ vanishes. 

\medskip

Then, using the relation $\widehat{N}{\widehat K}+\widehat D_e\widehat{N}{}^e=\tfrac12\,{\widehat \gamma}{}^{ab}\mathscr{L}_\rho{\widehat \gamma}{}_{ab}=\mathscr{L}_\rho\ln\sqrt{\widehat \gamma\,}$, together with the vanishing of the covariant divergence  of ${\widehat D}_a{\phi}{}^a$, we get
\begin{align}\label{eq: varJ3}
	{\phi}{}^a \mathscr{L}_{\rho}{ \rm \bf k}_a = {} & \widehat D_a \Big(\big[\widehat{N}{}^a({\phi}{}^e{ \rm \bf k}_e)-\widehat{N}\,\interior{\rm\bf K}{}^a{}_{e}{\phi}{}^e\big]+\widehat{N}\,\big[\tfrac12\,{\rm\bf K} +\boldsymbol\kappa\big]\,{\phi}{}^a \Big) - \widehat D_e {\phi}{}^a \Big(\widehat{N}{}^e { \rm \bf k}_a-\widehat{N}\,\interior{\rm\bf K}{}^e{}_{a}\Big) \nonumber \\ & - \big(\mathscr{L}_\rho\ln\sqrt{\widehat \gamma\,}\big)\,({\phi}{}^a{ \rm \bf k}_a) + 8\pi\widehat{N}\,T_{ef}n^e\widehat{\gamma}{}^{f}{}_{a} \,.
\end{align}
Combining then \eqref{eq: varJ2} and \eqref{eq: varJ3}, and using the vanishing of the integral of the total covariant divergences in \eqref{eq: varJ3} on the topological two-spheres $\mathscr{S}_\rho$, we get 
\begin{align}\label{eq: varJ4}
	\mathscr{L}_{\rho} J[\phi] =  {}&  - (8\pi)^{-1}\int_{\mathscr{S}_\rho}\Big[ \big(\mathscr{L}_{\rho}\,\ln\big[\underline{\sqrt{{\widehat{\gamma}}/{\interior{\gamma}}}}_{\,[{\tiny{\interior{\phi}}}]}/\sqrt{\widehat \gamma\,}\big]\big)\,({\phi}{}^a{ \rm \bf k}_a) +  \widehat D_a {\phi}{}_b \big(\widehat{N}\,\interior{\rm\bf K}{}^{ab}-\widehat{N}{}^a { \rm \bf k}^b\big)\,\Big]\, \widehat{{\boldsymbol{\epsilon}}} \nonumber \\ & - \int_{\mathscr{S}_\rho} \widehat{N}\,T_{ab}n^a{\phi}{}^b \, \widehat{{\boldsymbol{\epsilon}}} \,.
\end{align}
Integrating the last relation, with respect to $\rho$, on the part of $\Sigma$ between the level sets $\mathscr{S}_{{\rho}_1}$ and $\mathscr{S}_{{\rho}_2}$, denoted by $\Sigma_{1\rightarrow 2}$, we get
\begin{align}\label{eq: varJ5}
	J[\phi]\vert_{\mathscr{S}_{{\rho}_2}} {}& - J[\phi]\vert_{\mathscr{S}_{{\rho}_1}} = - \int_{\Sigma_{1\rightarrow 2}} T_{ab}n^a{\phi}{}^b \,  {\boldsymbol{\epsilon}}_{{}_{\Sigma}} \nonumber \\ & \hskip-0.6cm - (8\pi)^{-1}\int_{\Sigma_{1\rightarrow 2}} \widehat{N}^{-1}\,\Big[ \big(\mathscr{L}_{\rho}\,\ln\big[\underline{\sqrt{{\widehat{\gamma}}/{\interior{\gamma}}}}_{\,[{\tiny{\interior{\phi}}}]}/\sqrt{\widehat \gamma\,}\big]\big)\,({\phi}{}^a{ \rm \bf k}_a) +  \widehat D_a {\phi}{}_b \big(\widehat{N}\,\interior{\rm\bf K}{}^{ab}-\widehat{N}{}^a { \rm \bf k}^b\big)\,\Big]\, {\boldsymbol{\epsilon}}_{{}_{\Sigma}} \,,
\end{align}
where  ${\boldsymbol{\epsilon}}_{{}_{\Sigma}} = \widehat{N}\,d\rho\wedge\widehat{{\boldsymbol{\epsilon}}} $ denotes the volume element on $\Sigma$.  

\medskip

We conclude by emphasizing again that the balance relations \eqref{eq: varJ4} and \eqref{eq: varJ5} always hold regardless of the choice made for spherical coordinates $(\vartheta,\varphi)$ on $\mathscr{S}_{{\rho}_0}$. Note also that by virtue of the discussion in the last paragraph of section \ref{sec: axial-fields}, a spin-angular momentum vector field, denoted by  $\vec{J}\,[\interior{\phi},\rho]$, can be assigned to the balance relations. Note, however, that only the magnitude of these vector fields varies from sphere to sphere, since the spherical reference system $(\vartheta,\varphi)$ was chosen to be invariant under the action of the one-parameter group of diffeomorphisms induced by $\rho^a$.

\section{Final remarks}\label{sec: final-remarks}

It has been known for a long time that in the absence of well defined axial vector fields on generic Riemannian two-spheres no viable definition of quasilocal spin-angular momentum can be given. Our aims in this paper were twofold. First, we introduced a method that enable us to construct all the possible axial vector fields on generic Riemannian two-spheres. Then by restricting attention to the centre-of-mass  reference systems we also introduced a novel definition of the quasilocal spin-angular momentum that gets well defined for any three-dimensional spatial domain bounded by a sufficiently smooth topological two-sphere. By referring to the canonical embedding of the underlying unit sphere $(\mathscr{S},\interior{\gamma}{}_{ab})$ into $\mathbb{R}^3$ the quasilocal spin-angular momentum associated with the domain bounded by $\mathscr{S}$ can also be represented as a triple vector in the three-dimensional Euclidean space.

\medskip

We have also derived balance relations for quasilocal angular momentum expressions, with respect to some suitably chosen axial vector fields, on spacelike hypersurfaces foliated by topological two-spheres. These balance relations will allow to characterize the angular momentum transport in dynamical processes. It seems plausible that these balance relations can be extended to include hypersurfaces of arbitrary causal character. We also expect that these balance relations will prove to be essential monitoring tools in numerical investigations of highly dynamical processes of coupled gravity-matter systems in generic spacetimes without symmetries. 
 
\medskip 

Not surprisingly, a new construction raises many new issues. It remains to be seen whether the expectations regarding the global angular momentum charges (see references \cite{Wald-Zoupas-2000,Ashetkar-2022}), as discussed in the last paragraphs of section \ref{sec: quasilocal angular momentum}, can be verified. In addition, there are several practical issues that need to be addressed. First, given a two-metric ${\widehat{\gamma}}_{ab}$ on $\mathscr{S}$, one must determine a centre-of-mass unit sphere reference system.  Once this is done, the maximum of the integral expressions used in \eqref{eq: def-spin-angular-momentum} must be determined. These and many other questions are left for future investigations. Nevertheless, we believe that progress in research areas such as dynamical black hole thermodynamics and various aspects of Penrose inequalities will be greatly stimulated by the novel constructions and results presented in this paper. 

%%%%%%%%%%%%%%%%%%%% ACKNOWLEDGMENTS %%%%%%%%%%%%%%%%%%%%%%%%%%%%%%%%%%%
\section*{Acknowledgments}
 
The author would like to thank Ingemar Bengtsson, Károly Csukás, and Gábor Tóth for their careful reading and helpful comments. Special thanks are due to Jeff Winicour for his critical comments, which led to significant improvements in the content of this paper. 
This project was supported in part by the Hungarian Scientific Research fund NKFIH Grant No.~K-142423. 

%%%%%%%%%%%%%%%%%%%% Data Availability Statement %%%%%%%%%%%%%%%%%%%%%%%%%%%%%%%%%%%
%\section*{Data Availability}

%Data sharing not applicable to this article as no datasets were generated or analyzed during the current study.

%%%%%%%%%%%%%%%%%%%% REFERENCES %%%%%%%%%%%%%%%%%%%%%%%%%%%%%%%%%%%%%%%%
%\section*{References}

\end{document}